\documentclass[sigconf]{acmart}

\usepackage{natbib}
\usepackage{graphicx}
\usepackage{multirow}
\usepackage{calc}
\usepackage[normalem]{ulem}
\usepackage{tabularray}
\usepackage{float}
\usepackage[subfigure]{tocloft}
\usepackage{subfigure}
\usepackage{color}
\usepackage{tablefootnote}
\usepackage{soul}
\usepackage{algorithmic}
\usepackage{hyperref}
\usepackage{balance} 
\usepackage[ruled, linesnumbered]{algorithm2e}
\newcommand{\fullname}{\emph{M}ultimodality \emph{I}nvariant \emph{L}earning for \emph{CO}ld-Start recommendation}
\newcommand{\shortname}{\emph{MILK}}

\newcommand{\revise}[1]{{\color{blue}[#1]}}

\AtBeginDocument{%
  \providecommand\BibTeX{{%
    \normalfont B\kern-0.5em{\scshape i\kern-0.25em b}\kern-0.8em\TeX}}}

\copyrightyear{2024}
\acmYear{2024}
\setcopyright{acmlicensed}\acmConference[SIGIR '24]{Proceedings of the 47th International ACM SIGIR Conference on Research and Development in Information Retrieval}{July 14--18, 2024}{Washington, DC, USA}
\acmBooktitle{Proceedings of the 47th International ACM SIGIR Conference on Research and Development in Information Retrieval (SIGIR '24), July 14--18, 2024, Washington, DC, USA}
\acmDOI{10.1145/3626772.3658596}
\acmISBN{979-8-4007-0431-4/24/07}

\begin{document}

\title{Multimodality Invariant Learning for \\Multimedia-Based New Item Recommendation}

\author{Haoyue Bai}
\affiliation{\institution{Hefei University of Technology}
\city{Hefei}
\country{China}}
\email{baihaoyue621@gmail.com}

\author{Le Wu}\authornote{Corresponding author.}
\affiliation{\institution{Hefei University of Technology}
\institution{Institute of Dataspace,\\  Hefei Comprehensive National Science Center}
\city{Hefei}
\country{China}}
\email{lewu.ustc@gmail.com}

\author{Min Hou}
\affiliation{\institution{Hefei University of Technology}
\city{Hefei}
\country{China}}
\email{hmhoumin@gmail.com}

\author{Miaomiao Cai}
\affiliation{\institution{Hefei University of Technology}
\city{Hefei}
\country{China}}
\email{cmm.hfut@gmail.com}

\author{Zhuangzhuang He}
\affiliation{\institution{Hefei University of Technology}
\city{Hefei}
\country{China}}
\email{hyicheng223@gmail.com}

\author{Yuyang Zhou}
\affiliation{\institution{Academy of Cyber}
\city{Beijing}
\country{China}}
\email{yzhou193@alumni.jh.edu}

\author{Richang Hong}
\affiliation{\institution{Hefei University of Technology}
\institution{Institute of Dataspace,\\  Hefei Comprehensive National Science Center}
\city{Hefei}
\country{China}}
\email{hongrc.hfut@gmail.com}

\author{Meng Wang}
\affiliation{\institution{Hefei University of Technology}
\institution{Institute of Artificial Intelligence,\\  Hefei Comprehensive National Science Center}
\city{Hefei}
\country{China}}
\email{eric.mengwang@gmail.com}
\renewcommand{\shortauthors}{Bai, et al.}


\begin{abstract}

Multimedia-based recommendation provides personalized item suggestions by learning the content preferences of users. With the proliferation of digital devices and APPs, a huge number of new items are created rapidly over time. How to quickly provide recommendations for new items at the inference time is challenging. What's worse, real-world items exhibit varying degrees of modality missing~(e.g., many short videos are uploaded without text descriptions). Though many efforts have been devoted to multimedia-based recommendations, they either could not deal with new multimedia items or assumed the modality completeness in the modeling process. 

In this paper, we highlight the necessity of tackling the modality missing issue for new item recommendation. We argue that users' inherent content preference is stable and better kept invariant to arbitrary modality missing environments. Therefore, we approach this problem from a novel perspective of invariant learning. However, how to construct environments from finite user behavior training data to generalize any modality missing is challenging.
To tackle this issue, we propose a novel \textbf{M}ultimodality \textbf{I}nvariant \textbf{L}earning re\textbf{C}ommendation~(a.k.a. \shortname~) framework.
Specifically, \shortname~ first designs a 
cross-modality alignment module to keep semantic consistency from pretrained multimedia item features. After that, \shortname~ designs multi-modal heterogeneous environments with cyclic mixup to augment training data, in order to mimic any modality missing for invariant user preference learning.
Extensive experiments on three real datasets verify the superiority of our proposed framework.
The code is available at \href{https://github.com/HaoyueBai98/MILK}{https://github.com/HaoyueBai98/MILK}.

\end{abstract}


\maketitle

\section{Introduction}

With the proliferation of digital devices and APPs, individuals are exposed to abundant multimedia content, such as e-commerce and short-video sharing applications. Multimedia-based recommender systems have become indispensable components of these online services, aiming at learning user preferences from multimedia content to facilitate personalized item suggestions~\cite{He2015VBPRVB, Chen2017ACF, GoRec}. The key idea of these models lies in better model item content representations from pretrained features, and then align users' content preference with the help of users' historical records.

In real-world scenarios, multimedia-based recommender systems encounter distinctive characteristics. First, \textit{a huge number of new items emerge rapidly over time}, particularly on the news and short-video sharing platforms. E.g., about 500 hours of video are uploaded to YouTube every minute\footnote{https://www.statista.com/}~\cite{lin2023temporally}. 
Unlike the old items encountered during training, new items lack users' behavior data and need to be quickly recommended to keep freshness. 
Most of previous works focused on designing sophisticated content representations from the complete multimedia content~\cite{Sedhain2014SocialCF, geng2015learning, Lian2018xDeepFMCE, Oord2013DeepCM}.
Others proposed to leverage interchange between users and items with a fused graph from multiple content channels ~\cite{Wei2019MMGCNMG, Wei2020GraphRefinedCN, Zhang2021MiningLS, MICRO}. E.g., researchers proposed to exploit the user-item interaction records to guide the representation learning of each modality for final recommendation fusion~\cite{Wei2019MMGCNMG}. 
Most of these graph-based models show better performance for old items compared to pure content representation techniques.
When faced with new items, most models need to be retrained on new items, which could not satisfy the quick adaption to new items at the inference stage. Second, 
\textit{real-world items exhibit varying degrees of modality missingness}. For example, in short video-sharing platforms, authors may omit introductions for freshly uploaded videos, leading to a dearth of textual modality. Some new videos may intentionally lack audio features due to stylistic choices~\cite{DEMELO2020106179}. 
Nearly all previous works relied on modality completeness for recommendation or simply tackled this issue with preprocessing techniques to impute missing modalities. Therefore, the recommendation performance is hindered by the inferior preprocessing quality.  In summary, how to 
tackle the new item and the modality missing issues in multimedia recommendation is important and has not been well studied before.

\begin{figure}[t]
\centering
\includegraphics[width = 0.48\textwidth]{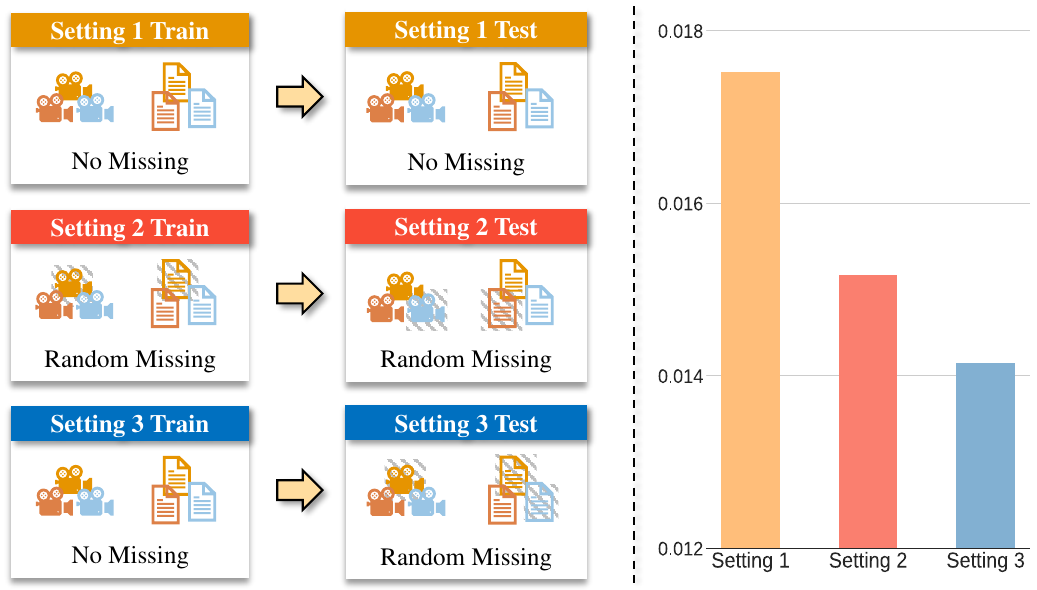}
    \caption{\small{New Item Recommendation with Missing Modalities}}
\label{fig:intro}
\end{figure}

In this paper, we study the problem of multimedia-based new item recommendation.  We provide an intuitive experiment to demonstrate the challenges of this problem. As shown in Figure~\ref{fig:intro}, we assess the performance of a classical multimedia-based recommendation model DUIF~\cite{geng2015learning} under different settings in Amazon Baby dataset~\cite{He2016UpsAD, McAuley2015ImageBasedRO}.
All samples in the original dataset have two modalities. The experimental settings are as follows:
(1) Original dataset with no modality missing.
(2) For the training and testing sets, samples randomly lose one modality.
(3) Only samples in the testing set randomly lose one modality.
Features for missing modalities are filled with the mean value.
We can clearly observe a decline in model performance in \textit{Setting 2} and \textit{3}, indicating that modality missingness has a substantial detrimental effect on the new item recommendation.
It should be noticed that the performance in \textit{Setting 3} is much worse than \textit{Setting 2}, showing that when there is a discrepancy in modality missingness between the training and testing sets, the model's performance experiences a more significant decline.
It also demonstrates that simple data preprocessing and removing training samples with modality missing are ineffective, and may even exacerbate the situation.

Essentially, the difficulty is that there are multiple combinations with modality missing.  When faced with new items of incomplete patterns, the inference stage distribution changes compared to the training stage.  
In fact, users' inherent content preference is stable and better kept invariant with arbitraty modality missing. Therefore, an idea recommendation model is encouraged to  predict each user's preference as invariant as possible. 

To achieve this goal, we draw inspiration from invariant learning~\cite{Arjovsky2019InvariantRM, Liu2021HeterogeneousRM,10.1145/3539618.3591729}, which can achieve guaranteed performance under distribution shifts and received great attention in recent years. Invariant learning models correlations invariant across different training environments, where environments are variables that should not affect the prediction.  Analogously, we want to learn users' inherent content preference that is invariant to any modality missing. However, implementing this analogy is challenging. 
As users' interaction records are limited, 
how to construct environments to generalize any modality missing is a challenge.


In this work, we propose a novel \textbf{M}ultimodality \textbf{I}nvariant \textbf{L}earning re\textbf{C}ommendation~(a.k.a. \shortname) framework for multimedia-based new item recommendation.
The main idea of \shortname~ is to encourage the users' inherent content preference stable and kept invariant to arbitrary modality missing scenarios for any new item recommendation.
The \shortname~ consists of two modules, the cross-modality alignment module for better item representation learning and the cross-environment invariance module for invariant preference prediction. 
Specifically, in the cross-modality alignment module, we devise alignment functions that allow one modality to capture content signals from other modalities. This module ensures that the absence of a specific modality does not hinder the extraction of features from other modalities.
In the cross-environment invariance module, we design cyclic mixup to create multi-modal heterogeneous environments and employ invariant learning to enhance the model’s generalization capability.
In cyclic mixup, we use the Dirichlet distribution to create diverse environments, allowing for imbalanced modality proportions and comprehensive consideration of each modality.

Our contributions are summarized as:
\begin{itemize}
\item We emphasize the significance of addressing the modality missing issue in multimedia-based new item recommendations. We argue that users' underlying preference is invariant to arbitrary modality missing environments, and tackle this problem from an invariant learning perspective.


\item We propose \shortname~ for the challenging problem. We design a novel cyclic mixup method, which constructs heterogeneous environments with different modal information proportions, thereby adapting invariant learning with continuous values and augmenting limited training data for invariant user preference modeling.

\item Extensive experiments on three real-world datasets demonstrate the superiority and effectiveness of \shortname~ in multimedia-based new item recommendation.
\end{itemize}

\section{Related Work}
\subsection{Multimedia-Based Recommendation}
The recommendation task aims to provide personalized recommendations to users\cite{Wu2021ASO, Wu2020DiffNetAN, Wu2021LearningFR}.
Multimedia-based recommendations utilize multimodal contents~(e.g., text, image, audio) of items to assist with the recommendation task.
The information-rich multimodal content greatly improves item characteristics and user preference modeling.
Early works incorporate the item visual features as side information into the models~\cite{He2015VBPRVB, Chen2017ACF, Kang2017VisuallyAwareFR}. 
For example, VBPR~\cite{He2015VBPRVB} adds a visual representation of items based on matrix factorization. 
Subsequently, some researchers utilize graph neural networks to perform embedding propagation on interaction graphs with different modality data, thereby capturing user preferences on different modalities~\cite{lin2023temporally, Wei2019MMGCNMG, Wei2020GraphRefinedCN, Li2020HierarchicalFG, Zhou2022ATO}.
For instance, MMGCN~\cite{Wei2019MMGCNMG} conducts graph convolutional operations on a modal-specific graph and captures the modal-specific user preference.
Although these works play important roles in exploring the use of multimodal information, most of them are transductive and rely heavily on CF information. 
Thus, they cannot flexibly deal with the constantly emerging new items.


To address this challenge, some works manage to enable models to make new item recommendations. 
These methods do not rely on CF information and have the ability to make recommendations for new items directly using multimedia features.
    Hybrid recommendation methods combine CF signals and multimedia information in the training stage to obtain hybrid preference representations~\cite{Zhu2020RecommendationFN, GoRec, Chen2022GenerativeAF, Wang2021PrivilegedGD, Zhou2023ContrastiveCF}. 
For example, GoRec~\cite{GoRec} directly models the distribution of pre-trained preference representations to generate representations for new items guided by multimedia features.
Content-based recommendation methods focus solely on item modeling based on multimedia features~\cite{Sedhain2014SocialCF, geng2015learning, Lian2018xDeepFMCE, Zhang2021MiningLS, MICRO}. For example, DUIF~\cite{geng2015learning} generates item representations using multimedia features and directly models user preferences for these features, enabling recommendations for new items based solely on their multimedia information. MICRO~\cite{MICRO} constructs and fuses multiple item-item relation graphs to explicitly mine the semantic information between items.
However, these methods often assume that the data is consistently complete and of high quality, a condition rarely met in the real world. In this paper, we address new item recommendations in scenarios where modalities are missing and introduce a more practical model to tackle this challenge.

\subsection{Invariant Learning for Recommendation}
Invariant learning~(IL)~\cite{Arjovsky2019InvariantRM,peters2016causal,chen2022does} improves the robustness of models to distribution shifts. IL is based on the assumption that the causal mechanism keeps invariant across various environments. 
By penalizing the variance of model prediction across environments, models are then encouraged to capture the causal mechanism instead of spurious correlations.
The common process of IL is to first split the training data into groups~(i.e., environments), here the groups need to reflect spurious correlations. Then, by ensuring consistent performances across different environments, the purpose of learning invariant representations is achieved.
The environment assignments play an important role in IL. 
Early works~\cite{Arjovsky2019InvariantRM} assume the environment labels are given in the dataset. Recently, IL has been introduced to the scenario where environment labels are unknown~\cite{creager2021environment,teney2021unshuffling}. These methods utilize prior knowledge of spurious correlations to split the training data.
For example, Teney \textit{et al.}~\cite{teney2021unshuffling} cluster training samples with their predefined spurious features. EIIL~\cite{creager2021environment} splits training data into the majority/minority sets on which the spurious feature conditioned label distribution varies maximally.
IL has been introduced into recommender systems for building more trustworthy models nowadays.
InvPref~\cite{Wang2022InvariantPL} estimates heterogeneous environments corresponding to different types of latent bias and uses IL to handle different unknown data biases in a unified framework.
InvRL~\cite{Du2022InvariantRL} utilizes environments to reflect the spurious correlations and then learns invariant representations to make a consistent prediction of user-item interaction across various environments.
In our work, we are inspired by IL, making the users' inherent content preference stable and kept invariant to arbitrary modality missing. We present a novel cyclic mixup
heterogeneous environments construction method. Cyclic mixup mimics infinite real-world scenarios with finite training samples.

\section{Problem Formulation}
\subsection{New Item Recommendation}
Let $\mathcal{U}$ and $\mathcal{V}$ denote the sets of users and items.
Since implicit feedback is very common, we use $\mathbf{R} \in \mathbb{R}^{|\mathcal{U}| \times |\mathcal{V}|}$ to denote the user-item interaction matrix, $r_{i j}=1$ if user $i$ has interacted with item $j$, otherwise $r_{i j}=0$.
Beyond the interaction signal, the multimodal features of items are extracted from their content, such as the visual, textual, acoustic modalities, and so on. We can generate the multimodality features into representations via generic feature extractors.
For an item $j \in \mathcal{V}$, we denote its feature vector as $\mathbf{x}_{j}\in\mathbb{R}^{M \times {d_x}}$, where $M$ is the number of modalities and $d_x$ is the dimension of the vector.
Taking the user ID $i$ and multimodality representation $\mathbf{x}_{j}$ of item $j$ as input, the recommendation model $\mathcal{F}_\Phi$ aims to infer the probability of user $i$ will interact with item $j$.
Optimizing $\mathcal{F}_\Phi$ is to minimize the loss function $\mathcal{L}$~(e.g., the BPR loss~\cite{Rendle2009BPRBP}) given the observed interactions:
\begin{equation}
\Phi^* = \mathop{\arg\min}\limits_{\Phi}\mathbb{E}_{(i,\mathbf{x}_{j})\sim\mathbf{P}_{train}}\mathcal{L}(\mathcal{F}_\Phi(i, \mathbf{x}_{j});\mathbf{R}).
\label{eq:opt_train}
\end{equation}
The expectation is calculated in the training data distribution $\mathbf{P}_{train}$. In the inference stage, we aim for the new item recommendation model $\mathcal{F}_{\Phi^*}$ can perform well on the new items set $\mathcal{V}_{\text{new}}$ without any historical interactions.

\subsection{Modality Missing Issue}
The new item recommendation models effectively infer the probability $\hat{y}_{ij}$ of user $i$ will interact with new item $j_{new}$ under the modality completeness assumption. Real-world scenarios often deviate from this idealized setting. Many samples exhibit varying degrees of modality missingness. 
We expect the recommendation model to remain effective in this scenario. 

We denote the modality representation of items' as $\mathbf{X}\in\mathbb{R}^{|V|\times M \times {d_{x}}}$. 
We use $\mathbf{A}\in\mathbb{R}^{|V|\times M}$ to denote the item modality indication matrix, where $\mathbf{a}_{jm}=1$ if the $m^\text{th}$ modalities of item $j$ is available, and otherwise $\mathbf{a}_{jm}=0$.
During the training stage, input data $(i,\mathbf{x}_{j})$ can be considered to be drawn from a joint distribution :
\begin{equation}
(i, \mathbf{x}_{j})\sim\mathbf{P}_{train}(i, \mathbb{I}(\mathbf{X}_j, \mathbf{A}_j)).
\end{equation}
$\mathbb{I}(\mathbf{X}_j, \mathbf{A}_j)$ represents some modality in $\mathbf{X}_j$ is missing indicated by $\mathbf{A}_j$.
In the inference stage, the unavailability of certain modalities in a new item leads to data being drawn from another joint distribution:
\begin{equation}
(i, \mathbf{x}_{j_\text{new}})\sim\mathbf{P}_{test}(i, \mathbb{I}(\mathbf{X}_\text{new}, \mathbf{A}_{j_\text{new}})).
\end{equation}
The multimodal data in the training set is typically complete, while the missingness of multimodal data in the testing phase is unknown, namely $\mathbf{A} \neq \mathbf{A_\text{new}}$.
The distribution shift between $\mathbf{P}_{test}$ and $\mathbf{P}_{train}$ challenges the performance of new item recommendation models built on the empirical risk minimization~(ERM) in the training set.

Our goal is to develop an optimal new item recommendation model capable of generalizing well to multimedia features drawn from the test distribution $\mathbf{P}_{test}$, where $\mathbf{P}_{test} \neq \mathbf{P}_{train}$. The optimization objective is to minimize the loss $\mathcal{L}(\mathcal{F}(i, \mathbf{x}_{j});\mathbf{R})$ with respect to the model's parameters $\Phi$, where the expectation is taken over the test data distribution $\mathbf{P}_{test}$. Note that $\mathbf{P}_{test}$ is unknown during the training stage.
\begin{equation}
\mathop{\arg\min}\limits_{\Phi}\mathbb{E}_{(i,\mathbf{x}_{j})\sim\mathbf{P}_{test}}\mathcal{L}(\mathcal{F}_\Phi(i, \mathbf{x}_{j});\mathbf{R}).
\label{eq:opt_test}
\end{equation}

\section{the proposed \shortname~ Framework}

\subsection{Overview of \shortname}
\begin{figure*}[ht]
\centering
\includegraphics[width = 1\textwidth]{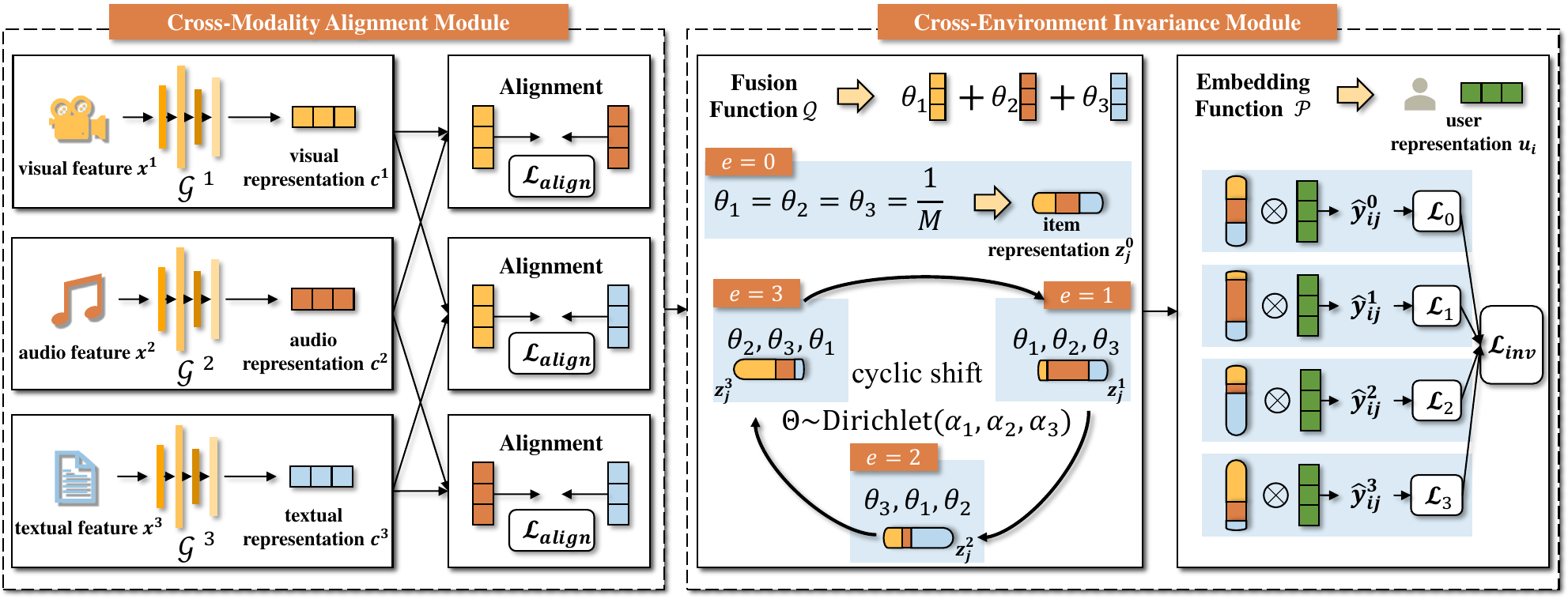}
\caption{\small{Model overview.
\shortname~ is consisted of Cross-Modality Alignment Module~(CMAM) and Cross-Environment Invariant Module~(CEIM).
CMAM obtains the modality representations $\mathbf{c}^m$ through independent feature extractors $\mathcal{G}^m$ and then imposes alignment between any two modalities.
CEIM converts user ID into user representation by embedding function $\mathcal{P}$ and generates item representations through fusion functions $\mathcal{Q}$.
CEIM generates multiple sets of weights as heterogeneous environments through cyclic mixup and aggregates multi-modal representations into item representations $\mathbf{z}^{e}_{j}$ in each environment $e$. 
Finally, CEIM optimizes the model under the invariant learning paradigm. }}
\label{fig:training}
\end{figure*}



As illustrated in Figure \ref{fig:training}, we present the overall framework of our proposed \shortname~. Essentially, \shortname~aims to encourage stable user preference prediction for new items, which is guaranteed by invariant preference learning under missing modality scenarios. To achieve this goal, \shortname~consists of two elaborated modules: Cross-Modality Alignment Module~(CMAM) and Cross-Environment Invariance Module~(CEIM). 

Specifically, CMAM aims to learn an informative multimodal representation under potential modality missing scenarios. We implement CMAM by narrowing each modality representation, thus each modality representation can supplemented from other modality features, to tackle the insufficient multimodal representation issue under missing modality. After multimodal representation alignment, we execute CEIM as follows: heterogeneous environment construction and invariant user preference learning in various environments. Particularly, we devise a flexible and unique environment construction based on a cyclic modality mixup strategy, enabling adaptation to varying proportions of missing modalities in test data. Given the constructed environments, we conduct stable user preference learning based invariant risk minimization~(IRM) principle. Next, we introduce each module in detail.

\subsection{Cross-Modality Alignment Module}
\noindent \textbf{Modal-Specific Extractors.}
Most existing works~\cite{Wei2021ContrastiveLF, Zhu2020RecommendationFN} concatenate multi-modal features and then convert them into item representation through an only extractor. 
This means that the absence of any modality has an impact on the whole extractor, which in turn adversely affects the extraction of other modal information.

CMAM uses independent extractors to guarantee stable modality feature extraction, which means each extractor is exclusively related to a specific modality.
Such an approach ensures the accuracy of the underlying understanding of existing modal features.
The process is as follows:
\begin{equation}
    \mathbf{c}^{m}_{j}=\mathcal{G}^{m}(\mathbf{x}^m_j)=\mathbf{W}^m\mathbf{x}^m_j+\mathbf{b}^m,
    \label{eq:extraction}
\end{equation}
where $\mathbf{x}^m_j$ denote the $m^{th}$ modalitys' original feature of item $j$, $\mathcal{G}^{m}$ denote the representation generator of $m^{th}$ modality and $\mathbf{W}^m$ and $\mathbf{b}^m$ are used to parameterize this process.

\noindent \textbf{Cross-Modality Alignment.} 
While the modal-specific extractors ensure the stable extraction of modal information, they also increase the risk that different modal information is mapped into different representation spaces.
We use alignment across modalities to guarantee the semantic consistency between representations of different modalities.
At the same time, the alignment across modalities makes the information between the modalities be transferred to each other. 
When a certain modality is unavailable, other modalities can provide a supplement. 
In \shortname, we achieve this goal by optimizing the following alignment objective:
\begin{equation}
    \mathcal{L}_{align}=\sum_{j=1}^{|\mathcal{V}|}\sum_{m=1}^{M-1}\sum_{m'=m+1}^M{{\mathbf{a}_{jm}\mathbf{a}_{jm'}\Vert\mathbf{c}^{m}_{j}-\mathbf{c}^{m'}_{j}\Vert}^2}.
    \label{eq:alignment}
\end{equation}


\subsection{Cross-Environment Invariant Module}
\noindent\textbf{Embedding and Fusion.}
CEIM obtains user representation and item representation by embedding layer  $\mathcal{P}$ and fusion function $\mathcal{Q}$.
For users, CEIM directly uses an embedding layer to convert the ID of user $i$ into a user representation $\mathbf{u}_{i}$.
Such an embedding layer can directly capture the user's interest from their interaction records.
For items, CMAM already generates $M$ representations $\mathbf{c}_j^m$ for each item.
Then CEIM fuse these representations by the function $\mathcal{Q}$ to obtain the representation of item $j$:
\begin{equation}
    \mathbf{z}_{j}=\mathcal{Q}{(\mathbf{c}_j^1,\ldots,\mathbf{c}_j^M)},
    \label{eq:fusion}
\end{equation}
In order to display and control the proportion of modal information more intuitively, we use weighted summation as the fusion function.
We first determine the weights $\Theta=\{\theta^1,\ldots,\theta^M\}$ and fuse the multimedia representations as follows:
\begin{equation}
    \mathbf{z}_{j}=\theta^1\times\mathbf{c}_j^1+\ldots+\theta^M\times\mathbf{c}_j^M.
    \label{eq: weighted sum}
\end{equation}

\noindent\textbf{Heterogeneous Environment Construction.}
To enable the model the ability to consistently perform well in complex modal missing scenarios, we assume the existence of multiple heterogeneous environments, each with multimedia features drawn from a different distribution. 
In our context, the environments should simulate the different modality missingness situations, i.e., let $(i, \mathbf{x}_{j})_{e}\sim\mathbf{P}_{train_{e}}(i, \mathbb{I}(\mathbf{X}_j, \mathbf{A}_{j_{e}}))$ indicates the training data belongs to environment $e$, $\forall e \neq e', \mathbf{P}_{train_{e}} \neq \mathbf{P}_{train_{e'}}$.
We encourage the model to maintain good performance and eventually learn users’ inherent content preferences across heterogeneous environments.
We naturally simulate and control the heterogeneity of the environment by adjusting the weights $\Theta$.

Fusing different modality representations using equal weights is the basic strategy.
In the testing phase, this fusion strategy will be used since no clear judgment can be made about the quality of information of a certain modality. 
We take equal weights $\Theta_{0}=\{\theta_{0}^1,\ldots,\theta_0^M\}=\{\frac{1}{M}\}^M$ into account as an environment during the training phase to ensure that this strategy is always exposed to the model, which ensures the basic performance of the model~\cite{Pinto2022RegMixupMA}.

Then, we construct heterogeneous environments by adjusting the weight.
We expect these environments to have some important properties:
(1) \textit{Unbalance Proportion}:
On the one hand, all modalities should be included in each environment, otherwise, the model will be encouraged to ignore modality, resulting in an overall decrease in performance. 
On the other hand, the proportion of modalities should be unbalanced to simulate the complex real situation.
(2) \textit{Full Modality Consideration}:  
Different environments should be dominated by different modalities, and each modality should play a major role in some environments. This guarantees that the model does not establish too strong associations with specific modalities and that the information of each modality is fully learned.
(3) \textit{Diversity for Generalization}: 
Enough environments should be simulated to improve the model's ability. 
The model should be exposed to a large number of heterogeneous environments.
If the number of environments is small, it may lead to spurious associations between the model and the limited pattern.

Inspired by mixup~\cite{Zhang2017mixupBE, Verma2018ManifoldMB}, a data augmentation method that mixes original samples to generate new samples, we propose a novel cyclic mixup method to construct environments satisfying the above properties.
Specifically, we first determine a Dirichlet distribution and then sample a set of weights  $\Theta_1=\{\theta_{1}^1,\ldots,\theta_1^M\}$ from this distribution:
\begin{equation}
    \Theta_1\sim\mathit{Dirichlet}(\alpha_1,\ldots,\alpha_M),
    \label{eq:deli}
\end{equation}
where ${\alpha_1,\ldots,\alpha_M}$ is used to adjust the Dirichlet distribution. 
Since the Dirichlet distribution represents a probabilistic simplex in M-dimensional space, there is no need for additional normalization of the weights and we can extend this method to more scenarios with any number of modalities.

We then generate the other weights by cyclic shift:
\begin{equation}
    \Theta_m=\mathit{circle\_shift}(\Theta_{m-1}),
    \label{eq:cs}
\end{equation}
where $\Theta_m$ denotes the resulting weight after $m-1$ cyclic shift.
Each time through the shift, we move all the weights in $\Theta_{m-1}$ back one bit and place the last weight at the front.
The weight changes for performing one cyclic shift are as follows:
\begin{equation}
    \{\theta^1,\theta^2,\ldots,\theta^m,\ldots,\theta^M\}\longrightarrow\{\theta^M,\theta^1,\ldots,\theta^{m-1},\ldots,\theta^{M-1}\}.
    \label{eq:cs_e}
\end{equation}

We can construct $M$ environments based on $\Theta_1$ by $M\mbox {-}1$ cyclic shift. 
The weights sampled from the Dirichlet distribution guarantee the \textit{Unbalance Proportion} that the weights are nonzero and unequal for different modes.
Cyclic shift guarantees the property of \textit{Full Modality Consideration}, where the dominant mode is different in each environment, and each mode will dominate an environment. 
We perform this process at each iteration, which ensures the \textit{Diversity for Generalization} of the environment through randomness.

We denote environments by $e$, where $e=0$ is the environment with equal weights, and $e={1,\ldots, M}$ denotes the $M$ environments obtained by cyclic mixup. 
This fusion function uses each environment $e$'s weight to generate the item representation $\mathbf{z}^{e}_{j}$:
\begin{equation}
    \mathbf{z}^{e}_{j}=\mathcal{Q}(\mathbf{c}_j^1,\ldots,\mathbf{c}_j^M)=\theta_{e}^1\times\mathbf{c}_j^1+\ldots+\theta_{e}^M\times\mathbf{c}_j^M.
    \label{eq: weighted sum}
\end{equation}

\noindent\textbf{Invariant optimization.}
With the synergy of multiple modules, the model can predict the user's liking for the item based on the user ID and item multimedia features in each environment:
\begin{equation}
    \hat { y }^{e} _ { ij }=\mathcal{F}_\Phi(\mathbf{u}_i, \mathbf{z}^{e}_{j}),
    \label{eq:model}
\end{equation}
herein, $\mathcal{F}_\Phi$ estimates the probability that a user likes an item by the inner product operation.

We argue that users’ inherent content preference is stable and better kept invariant to arbitrary modality information distribution.
Hence, we encourage the model to maintain performance in heterogeneous environments.
To do this, we train the model using the invariant learning paradigm to learn an invariant preference prediction mechanism.
We use one of the common optimization objectives under the invariant learning paradigm as follows~\cite{Krueger2020OutofDistributionGV}:
\begin{equation}
    \mathcal{L}_{inv}=\mathbb{E}_{e\in\mathcal{E}}\mathcal{L}_e+\beta\mathbf{Var}_{e\in\mathcal{E}}(\mathcal{L}_e),
    \label{eq:irm}
\end{equation}
where the first term is the task-dependent loss used to guarantee the performance of the model on the target task.  The second term is the constraint over the loss variance across environments, which encourages the model to be stable across different environments.
In the context of our work, the environment $e$ is represented by different weights $\Theta_m$, and $\mathcal{L}_e$ is the average recommendation loss value inside the environment $e$, and the widely used BPR loss is adopted in this paper:
\begin{equation}
     \mathcal{L}_e = \sum _ { ( i, j , j' ) \in {\mathcal{U} \cup \mathcal{V}}} - \ln \sigma \left( \hat { y }^{e} _ { ij } - \hat { y }^{e} _ { ij' } \right),
\label{eq:bpr}
\end{equation}
where $j'$ is the negative sample sampled item of user $i$.

\subsection{Model Optimization and Inference}
In summary, our final optimization objective is:
\begin{equation}
    \mathcal{L}=\mathbb{E}_{e\in\mathcal{E}}\mathcal{L}_e+\beta\mathbf{Var}_{e\in\mathcal{E}}(\mathcal{L}_e)+\lambda\mathcal{L}_{align}+ \| \Phi \| ^ { 2 },
    \label{eq:opt}
\end{equation}
where $\lambda$ is a hyperparameter that controls the weight of the alignment loss, and $\Phi$ includes all model parameters.
We optimize the model parameters in the training stage by :
\begin{equation}
\Phi^* = \mathop{\arg\min}\limits_{\Phi}\mathbb{E}_{(i,\mathbf{x}_{j})\sim\mathbf{P}_{train}}\mathcal{L}(\mathcal{F}_\Phi(i, \mathbf{x}_{j});\mathbf{R}).
\label{eq:opt_train}
\end{equation}

In the inference stage, \shortname~ can directly apply to new items with varying degrees of modality missingness. When a new item $j_\text{new}$ appears, the multimedia feature $\mathbf{X}_{j_\text{new}}$ is first processed using mean imputation. Then we can predict the preference score of user $i$ to $j_{\text{new}}$ as:
\begin{equation}
    \hat { y } _ { ij_\text{new}}=\mathcal{F}_{\Phi^*}(\mathbf{u}_i, \frac{1}{M}\sum_{m=1}^{M}\mathcal{G}^{m}({\mathbf{x}}^m_{j_\text{new}})).
    \label{eq:infer}
\end{equation}

\section{Experiments}
In this section, we conduct extensive experiments on three real-world datasets, which aim to answer the following questions:
\begin{itemize}
    \item \textbf{Q1}: How does our model perform compared with state-of-the-art new item recommendation methods?
    \item \textbf{Q2}: How does our model improve new item recommendation performance with missing modalities?
    \item \textbf{Q3}: How do all modules in our model make positive effects on the performance?
\end{itemize}
\subsection{Experimental Settings}
\subsubsection{Datasets}
\noindent \textbf{Datasets Description.} 
We conduct experiments on three widely used real-world datasets, including (a) Amazon Baby~\cite{He2016UpsAD, McAuley2015ImageBasedRO}, (b) Amazon Clothing, Shoes, and Jewelry~\cite{He2016UpsAD, McAuley2015ImageBasedRO}, and (c) TikTok\footnote{http://ai-lab-challenge.bytedance.com/tce/vc/}. To simplify reading, they are called Baby, Clothing, and TikTok.
Baby and Clothing include both visual and textual modalities. 
TikTok is collected from the TikTok platform to log the viewed short videos of users. 
The multi-modal features are visual, acoustic, and title textual features of videos.
For a fair comparison, all models use the pretrained multi-modal features as input~\cite{Zhang2021MiningLS, wei2023multi}. 
The statistics of the pre-processed datasets are listed in Table \ref{tab: statistics}.

\noindent \textbf{New Item Setting.} 
We randomly select 20\% items and delete their historical interactions in the training process to simulate new items. 
Among them, we further divide half as validation and the remaining as test items. 
These items are entirely unseen in the training set.

\noindent \textbf{Modality Missing Setting.} 
On three datasets, we validate the effect of our model under two settings.
In the first Full Training Missing Test~(FTMT) setting, we assume that the quality of the training data can be guaranteed. 
We use the complete multimedia features for training in the training phase, and randomly select 50\% of the items in the testing phase, assuming that they randomly miss one modality feature. 
The missing modality is randomly selected and filled in using the mean imputation to pre-process (We use a simple padding way to keep the input data in a consistent format, and we provide detailed experiments on the imputation method in section~\ref{imputation}). 
In the second more realistic setting Missing Training Missing Test~(MTMT), we assume that there are also 30\% items in the training set that randomly miss one modality feature. 
\begin{table}
\small
\centering
\caption{\small{The statistics of datasets.}}
\label{tab: statistics}
\begin{tabular}{c|l|c|c|c} 
\hline
\multicolumn{2}{c|}{Dataset}             & Baby    & Clothing & TikTok   \\ 
\hline\hline
\multirow{4}{*}{Train} & \# Users        & 19442   & 39384    & 9319     \\
                       & \# Items        & 5640    & 18427    & 5368     \\
                       & \# Interactions & 128963  & 222759   & 55126    \\
                       & Density         & 0.118\% & 0.031\%  & 0.110\%  \\ 
\hline
\multirow{2}{*}{Val}   & \# Users        & 10342   & 19801    & 2380     \\
                       & \# Items        & 705     & 2303     & 671      \\ 
\hline
\multirow{2}{*}{Test}  & \# Users        & 10474   & 19858    & 2960     \\
                       & \# Items        & 705     & 2303     & 671      \\
\hline
\end{tabular}
\end{table}
\subsubsection{Evaluation Metrics} 
We select two metrics that are widely used in personalized recommender systems: Recall (Recall@K) and Normalized Discounted Cumulative Gain (NDCG@K).
The higher these two metrics are, the better the model is performing.

\subsubsection{Baselines} 

To verify the effectiveness of \shortname, we select multiple SOTA models suitable for task scenarios for comparison:
\begin{itemize}
    \item \textbf{DUIF}~\cite{geng2015learning} is a content-based method. It transforms heterogeneous user-content networks into homogeneous low-dimensional space for unified representation learning.
    \item \textbf{MICRO}~\cite{MICRO} is a hybrid method. It learns item-item relationships for each modality, and it utilizes contrastive learning for better item-level multimodal fusion.
    \item\textbf{DropoutNet}~\cite{Volkovs2017DropoutNetAC} is designed for new item recommendation. It randomly drops the partial preference representation in the training stage to simulate the new item scenario.
    \item\textbf{MTPR}~\cite{Du2020HowTL} address the new item multimedia recommendation issue by strategically replacing the preference representation with the all-zero vector in the training phase.
    \item\textbf{Heater}~\cite{Zhu2020RecommendationFN} is a new item recommendation method that uses the sum squared error loss to align CF signal and content representation.
    \item\textbf{CLCRec}~\cite{Wei2021ContrastiveLF} uses contrastive learning to constrain CF signal and content representation to maximize the mutual information between them.
    \item\textbf{CCFCRec}~\cite{Zhou2023ContrastiveCF} capitalizes on the co-occurrence collaborative signals in warm training data to alleviate the issue of blurry collaborative embeddings.
    \item\textbf{GAR}~\cite{Chen2022GenerativeAF} trains a generator and a recommender adversarially, generating the new item's representation which is similar to the old item's representation. 
    \item\textbf{GoRec}~\cite{GoRec} directly models the distribution of pre-trained preference representations to generate representations for new items guided by multimedia features.
\end{itemize}
We comprehensively selected multiple content-based recommendation methods and multimedia new-item recommendation methods as baselines. Note that some graph-based multimedia recommendation methods (e.g., MMGCN) are not compared because they require graph construction based on user-item interaction records, which makes it difficult be apply in our challenging scenario.

\subsubsection{Hyper-Parameter Settings}
We implement our \shortname~ and all baselines with Pytorch framework\footnote{https://pytorch.org/}.
The dimension of preference representation is fixed as 64.
The batch size is set to 2048. 
During training, we employ Adam~\cite{Kingma2014AdamAM} as the optimizer and set the learning rate at 0.001, the early stop strategy is employed to avoid over-fitting.
We carefully search the best parameter of $\beta$ and $\lambda$ and find \shortname~ achieves the best performance when $\beta=1000$ and $\lambda=0.05$ on Baby, $\beta=50$ and $\lambda=0.05$ on Clothing, and $\beta=50$ and $\lambda=0.5$ on TikTok dataset.
For the choice of the Dirichlet distribution when constructing the environment, we set $\alpha_{1}=\ldots=\alpha_{M}$ and search $\alpha_m$ from [0.01, 0.1, 1, 10, 100].
For all baselines, we search the parameters carefully for fair comparisons. 
We repeat all experiments 5 times and report the average results.

\subsection{Overall Comparisons (Q1)} \label{overall comparisons}

\begin{table*}
\caption{\small{Performance comparisons with different Top-K values different settings.}}
\small
\label{tab: overall}
\centering
\begin{tabular}{c|c|cc|cc|cc|cc|cc|cc} 
\hline
\multicolumn{2}{c|}{Settings}          & \multicolumn{6}{c|}{~Full Training Missing Test (FTMT)}                                                   & \multicolumn{6}{c}{~Missing Training Missing Test (MTMT)}                                                  \\ 
\hline
\multicolumn{2}{c|}{Datasets}          & \multicolumn{2}{c|}{Baby}         & \multicolumn{2}{c|}{Clothing}     & \multicolumn{2}{c|}{Tiktok}       & \multicolumn{2}{c|}{Baby}         & \multicolumn{2}{c|}{Clothing}     & \multicolumn{2}{c}{Tiktok}         \\ 
\hline
Metric                    & Models     & K=10            & K=20            & K=10            & K=20            & K=10            & K=20            & K=10            & K=20            & K=10            & K=20            & K=10            & K=20             \\ 
\hline\hline
\multirow{9}{*}{Recall@K} & DUIF       & 0.0217          & 0.0381          & 0.0189          & 0.0318          & \uline{0.1814}  & \uline{0.1921}  & 0.0172          & 0.0308          & 0.0188          & 0.0323          & 0.1706          & 0.1804           \\
                          & DropoutNet & 0.0178          & 0.0208          & 0.0155          & 0.0269          & 0.1221          & 0.1276          & 0.0130          & 0.0227          & 0.0150          & 0.0254          & 0.1011          & 0.1169           \\
                          & MTPR       & 0.0261          & 0.0328          & 0.0224          & 0.0371          & 0.1526          & 0.1615          & 0.0187          & 0.0325          & 0.0190          & 0.0317          & 0.1444          & 0.1661           \\
                          & Heater~    & 0.0303          & 0.0492          & 0.0338          & 0.0522          & 0.1260          & 0.1327          & 0.0259          & 0.0438          & 0.0308          & 0.0515          & 0.1148          & 0.1349           \\
                          & CLCRec     & 0.0247          & 0.0418          & 0.0281          & 0.0453          & 0.1529          & 0.1584          & 0.0190          & 0.0328          & 0.0265          & 0.0403          & 0.1228          & 0.1459           \\
                          & CCFCRec    & 0.0295          & 0.0464          & 0.0355          & 0.0545          & 0.1802          & 0.1920          & 0.0210          & 0.0355          & 0.0291          & 0.0453          & \uline{0.1715}  & \uline{0.1809}   \\
                          & GAR~       & 0.0307          & 0.0485          & 0.0352          & 0.0555          & 0.1486          & 0.1691          & 0.0271          & 0.0449          & 0.0319          & 0.0523          & 0.1292          & 0.1512           \\
                          & GoRec      & \uline{0.0335}  & \uline{0.0544}  & \uline{0.0429}  & \uline{0.0639}  & 0.1465          & 0.1605          & \uline{0.0282}  & \uline{0.0473}  & \uline{0.0430}  & \uline{0.0634}  & 0.1416          & 0.1582           \\
                          & Ours       & \textbf{0.0381} & \textbf{0.0628} & \textbf{0.0484} & \textbf{0.0716} & \textbf{0.1886} & \textbf{0.2002} & \textbf{0.0328} & \textbf{0.0517} & \textbf{0.0470} & \textbf{0.0684} & \textbf{0.1793} & \textbf{0.1884}  \\ 
\hline\hline
\multirow{9}{*}{NDCG@K}   & DUIF       & 0.0117          & 0.0163          & 0.0100          & 0.0137          & 0.1792          & \uline{0.1803}  & 0.0088          & 0.0125          & 0.0104          & 0.0140          & \uline{0.1682}  & \uline{0.1706}   \\
                          & DropoutNet & 0.0057          & 0.0071          & 0.0093          & 0.0123          & 0.0979          & 0.0956          & 0.0071          & 0.0098          & 0.0077          & 0.0100          & 0.0824          & 0.0860           \\
                          & MTPR       & 0.0156          & 0.0208          & 0.0125          & 0.0165          & 0.1271          & 0.1211          & 0.0100          & 0.0139          & 0.0106          & 0.0140          & 0.1177          & 0.1229           \\
                          & Heater~    & 0.0178          & 0.0222          & 0.0186          & 0.0237          & 0.1082          & 0.1016          & 0.0136          & 0.0185          & 0.0166          & 0.0223          & 0.0937          & 0.0975           \\
                          & CLCRec     & 0.0133          & 0.0181          & 0.0163          & 0.0210          & 0.1321          & 0.1249          & 0.0105          & 0.0143          & 0.0148          & 0.0185          & 0.1036          & 0.1080           \\
                          & CCFCRec    & 0.0159          & 0.0209          & 0.0207          & 0.0259          & \uline{0.1799}  & 0.1802          & 0.0114          & 0.0154          & 0.0163          & 0.0207          & 0.1567          & 0.1596           \\
                          & GAR~       & 0.0163          & 0.0210          & 0.0200          & 0.0255          & 0.1133          & 0.1173          & 0.0147          & 0.0196          & 0.0176          & 0.0232          & 0.0956          & 0.1004           \\
                          & GoRec      & \uline{0.0179}  & \uline{0.0249}  & \uline{0.0234}  & \uline{0.0292}  & 0.1096          & 0.1124          & \uline{0.0153}  & \uline{0.0206}  & \uline{0.0234}  & \uline{0.0300}  & 0.0952          & 0.0988           \\
                          & Ours       & \textbf{0.0215} & \textbf{0.0277} & \textbf{0.0279} & \textbf{0.0343} & \textbf{0.1830} & \textbf{0.1852} & \textbf{0.0182} & \textbf{0.0234} & \textbf{0.0270} & \textbf{0.0329} & \textbf{0.1774} & \textbf{0.1779}  \\
\hline
\end{tabular}
\end{table*}

As shown in Table \ref{tab: overall}, we compare our model with other baselines on three datasets. 
We have the following observations:

(1) On all three datasets, \shortname~ shows a significant improvement over all baselines in the FTMT setting.
Specifically, \shortname~ improves the strongest baseline $w.r.t$ NDCG@20 by 11.2\%, 14.0\%, and 2.8\% on Baby, Clothing, and TikTok dataset, respectively. 
Extensive empirical studies verify the effectiveness of the proposed \shortname~. We attribute this improvement to the two modules we designed, which greatly enhanced the generalization ability of the model.

(2) In the MTMT setting, we consider scenarios that are more realistic. The lack of partial item modality in the training phase puts forward higher requirements for the fitting and generalization ability of the model. \shortname~ still performs best on all three datasets. \shortname~ improves the strongest baseline $w.r.t$ NDCG@20 by 13.5\%, 9.6\%, and 4.3\% on Baby, Clothing, and TikTok dataset, respectively. The experimental results further validate the practicability of \shortname.

(3) DUIF, MICRO, and \shortname~ only model the characteristics of new items through multi-modal features, without introducing additional CF representation. Our model performs well on all settings and datasets, which fully demonstrates that our design is working. For instance, on Baby dataset \shortname~ improves DUIF $w.r.t$ NDCG@20 by 69.9\%, 87.2\%, in FTMT setting and MTMT setting.

\subsection{Improvements on Modality Missingness (Q2)} 

\subsubsection{Performance on Different Missing Scenarios} 
\begin{figure}[ht]
\centering
\includegraphics[width =0.35\textwidth]{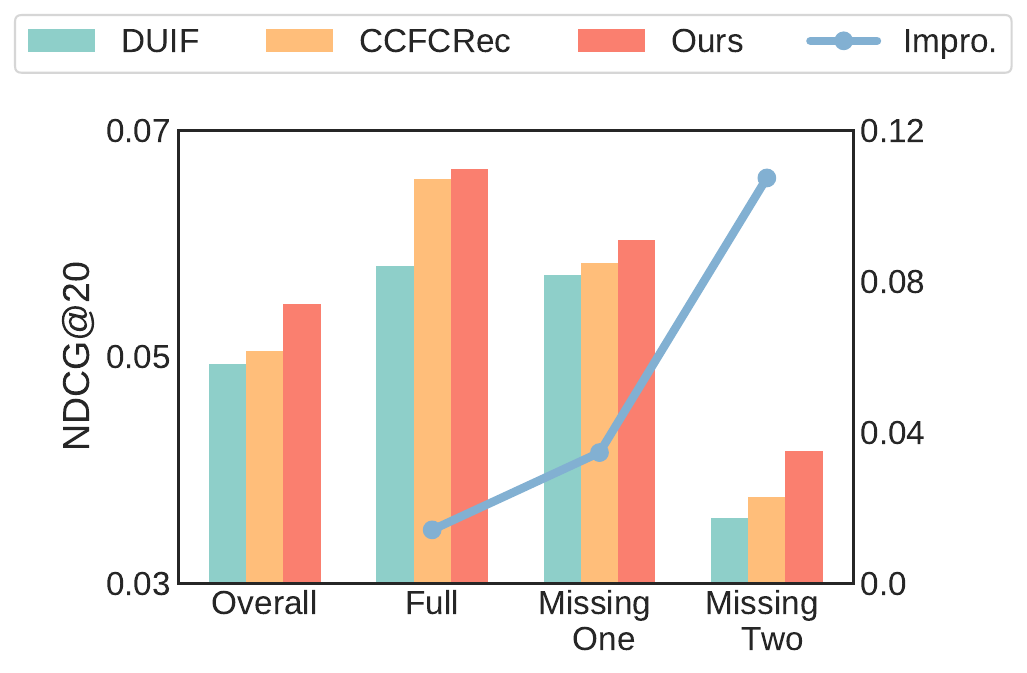}
\caption{\small{Performance on different missing scenarios.}}
\label{fig: worst}
\end{figure}
In this section, we further explore how \shortname~ improves performance when the modalities are missing.
The Tiktok dataset has richer modalities, based on which we construct different modality missing scenarios.
We use the full training and validation sets. 
In the test set, 10\% items are missing no modality, 45\% are randomly missing one modality and 45\% are randomly missing two modalities.
We report the performance of the model on all samples and on a specific group of items separately.
We select the top 2 baselines on the Tiktok dataset in Section \ref{overall comparisons} for comparison.
The bar charts in Figure \ref{fig: worst} show the results of the three models under different groups. \textit{Overall} represents the results on all test sets, \textit{full} represents the results on the item group with complete modality, \textit{missing one} and \textit{missing two} represent the item group with one and to modality missing, respectively. 
For ease of presentation, we show the numbers after two decimal places of NDCG@20.
The line chart demonstrates the percentage of our method improving the strongest baseline.
We can observe that the models outperform baselines on the whole test set. 
At the same time, we find that the improvement of the \shortname~ is more obvious in the item group with missing modalities, and the improvement of the group with missing two modalities is far more than the rest of the groups.
This shows that our model really improves the recommendation performance of the model on modal missing items, and has stronger generalization ability.

\subsubsection{Comparison with common data imputation methods.} \label{imputation} 
\begin{table}
\centering
\caption{Comparison with common data imputation methods.}
\label{tab: imputation}
\begin{tabular}{c|cc|cc} 
\hline
Datasets & \multicolumn{2}{c|}{Baby} & \multicolumn{2}{c}{Clothing}  \\ 
\hline
Metrics  & Recall@20 & NDCG@20       & Recall@20 & NDCG@20           \\ 
\hline\hline
Zero     & 0.0501    & 0.0215        & 0.0670     & 0.0322            \\
Mean     & 0.0525    & 0.0228        & 0.0665    & 0.0320             \\
Map      & 0.0507    & 0.0220         & 0.0619    & 0.0294            \\
CMAM     & 0.0585    & 0.0255        & 0.0682    & 0.0330             \\
CEIM     & 0.0609    & 0.0264        & 0.0686    & 0.0335            \\
Ours    & 0.0628    & 0.0277        & 0.0716    & 0.0343            \\
\hline
\end{tabular}
\end{table}

We further designed experiments to verify the effects of different data imputation operations in the inference stage. We conduct experiments on Baby and Clothing datasets in the FTMT setting and report the result in Table \ref{tab: imputation}. \textit{Zero} means that we delete the alignment part, heterogeneous environment construction and invariant learning part in ~\shortname, and fill the missing modalities in the test set with zeros. \textit{Mean} indicates padding with the mean value. \textit{Map} indicates we pre-train the mapping function between the two modalities. In the testing phase, for the missing image modality, we use the text\_to\_image mapping function to calculate image representation from the text representation. The missing text representation is computed from the image representation using the image\_to\_text mapping function. \textit{CMAM} and \textit{CEIM} indicate that we used only one module from ~\shortname.
The experimental results show that the preprocessing method does not improve the model performance in our problem scenario. Sometimes a well-designed imputation method can further degrade the performance of the model, such as the \textit{Mean} and \textit{Map} imputation on the Clothing dataset. In contrast, each component of ~\shortname~ can effectively improve the performance of the model in the absence of modalities.

\subsubsection{Impact of Constructed Environment} \label{env}
\begin{table}
\small
\centering
\caption{Different strategies for constructing environments.}
\renewcommand\arraystretch{0.9}
\label{tab:env}
\begin{tabular}{c|cc|cc} 
\hline
Dataset   & \multicolumn{2}{c|}{Baby} & \multicolumn{2}{c}{Clothing}  \\ 
\hline
Metric    & Recall@20 & NDCG@20       & Recall@20 & NDCG@20           \\ 
\hline\hline
ours      & 0.0628    & 0.0277        & 0.0716    & 0.0343            \\
w/o e=0   & 0.0433    & 0.0179        & 0.0685    & 0.0333            \\
w/o cs    & 0.0487    & 0.0205        & 0.0656    & 0.0319            \\
w/o random & 0.0555    & 0.0237        & 0.0643    & 0.0314            \\
\hline
\end{tabular}
\end{table}

\shortname~  constructs differentiated environments by adjusting modal weights and learns preference prediction mechanisms that are invariant across environments. 
We further explore the impact of different environmenta construction strategies.
\textit{w/o e=0} does not additionally consider the case with the same weights.
\textit{w/o cs} indicates that no cyclic shift is used to generate the weights, and weights are sampled from the Dirichlet distribution for each environment while keeping the number of environments.
\textit{w/o random} means that weights are sampled and cyclically shifted to construct the environment only at the beginning of model training, and weights are fixed afterward.
Table \ref{tab:env} shows that our method outperforms the three variants of the constructive environment method. Our strategy for constructing the environment is reasonable, necessary and effective.

\subsection{Detailed Model Analysis (Q3)} \label{Detailed Model Analysis}
\subsubsection{Ablation Study} \label{ablation study}
\begin{figure}[t]
\centering
\includegraphics[width =0.4\textwidth]{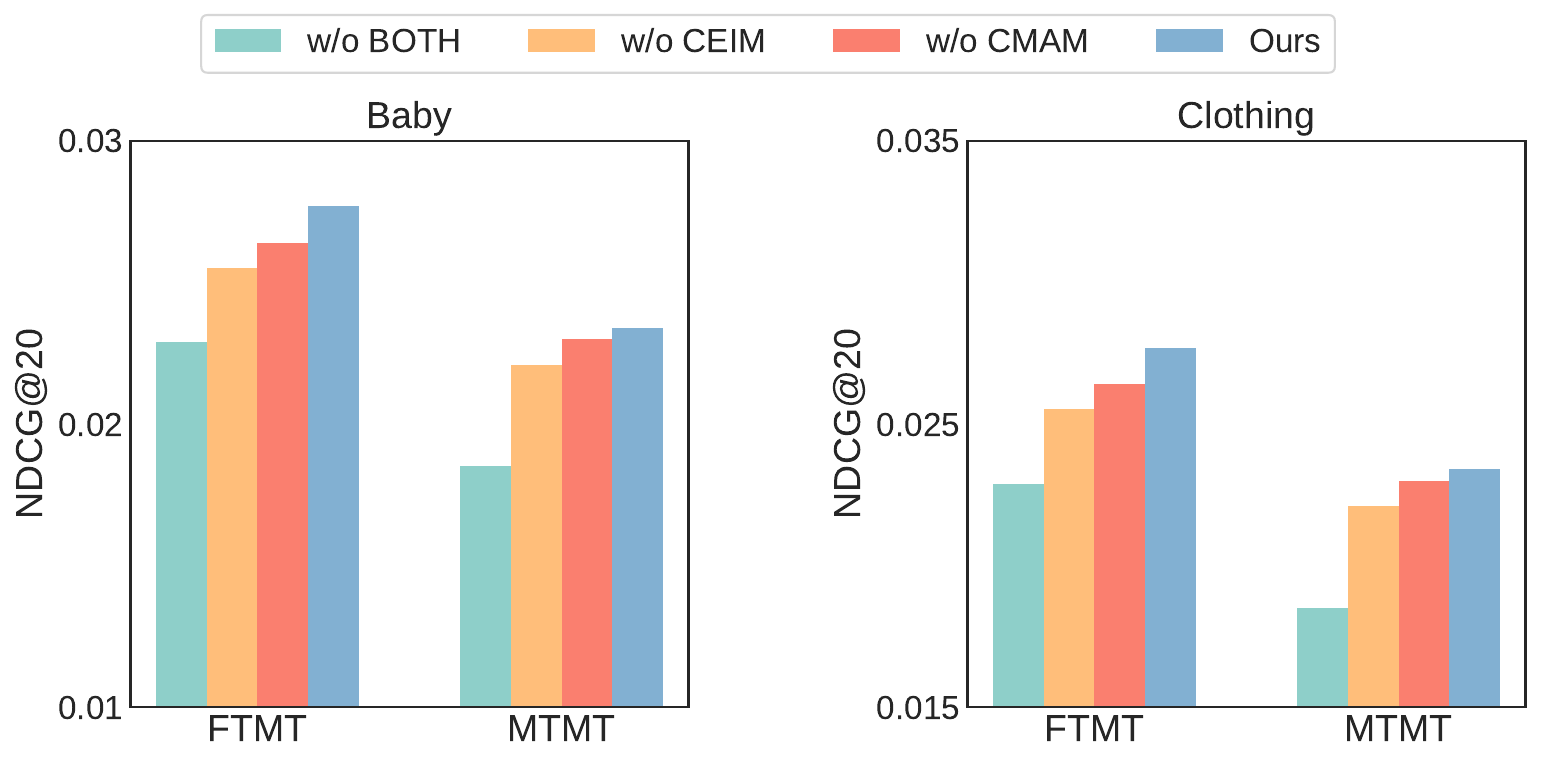}
\caption{\small{Ablation experiments on Baby and Clothing datasets.}}
\label{fig: ablation}
\end{figure}

To exploit the effectiveness of each component of the proposed \shortname~, we conduct the ablation study on different datasets. 
As shown in Figure \ref{fig: ablation}, we compare \shortname~ and corresponding variants on Top-20 recommendation performance. 
\textit{\shortname-w/o CMAM} denotes that the Cross-Modality alignment module is removed and the fusion is performed directly after extracting the modal representation.
\textit{\shortname-w/o CEIM} refers to training under the ERM paradigm. In this setting, we directly use the average strategy to fuse the representations of multiple modalities, and then calculate the propensity score with the user's representation to optimize the objective by BPR.
\textit{\shortname-w/o BOTH} means that we delete both modules at the same time. Under this setting, we did not perform any design for distribution inconsistency.
In Figure \ref{fig: ablation}, we observed that each component of the \shortname~ contributed to the final superior performance. 
The boost of either module for \textit{\shortname-w/o BOTH} is huge. For example, in the FTMT setting, CMAM and CEIM improve \textit{\shortname-w/o BOTH} $w.r.t$ NDCG@20 by 19.4\%, and 24.27\%  on the Baby dataset, respectively.

 \begin{figure} [t]
  \centering
      \subfigure[Baby]{
      \includegraphics[width=0.23\textwidth]{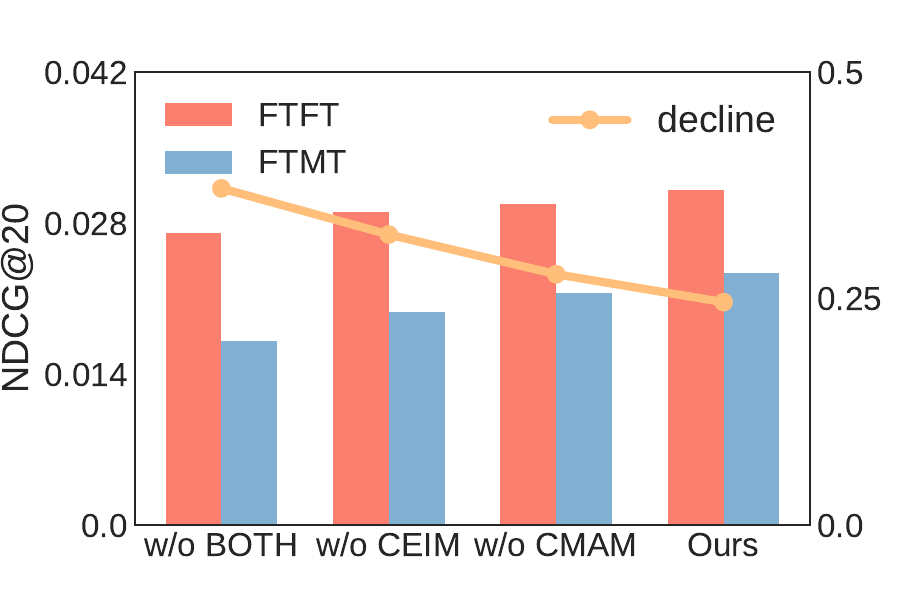}} 
      \subfigure[Clothing]{
      \includegraphics[width=0.23\textwidth]{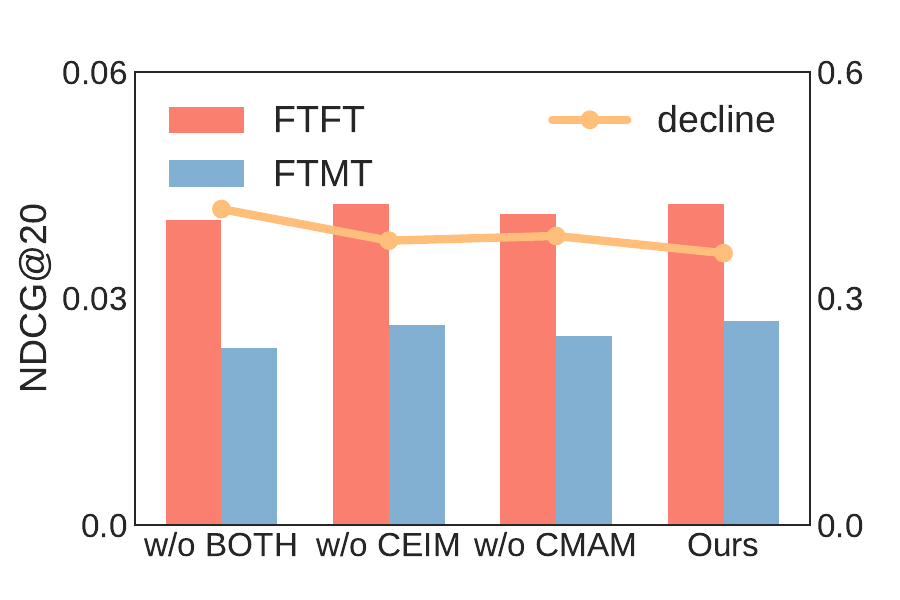}}
  \caption{\small{Effect of different modules on the robustness of the model.}}
  \label{fig: reduce}
\end{figure}
Besides, we focus more on the generalization ability of the model in missing modality scenarios. 
As shown in Figure \ref{fig: reduce}, we report the impact of different modules on the model in the full scenario (Full modality Training, Full modality Test (FTFT)) and missing modality scenario (FTMT).
The bar charts represent the performance of the model variants under different scenarios. 
We can clearly see that the performance of the model drops significantly in the missing modality scenario. 
The line chart visually shows the magnitude of performance degradation for different model variants in the two Settings. 
The addition of the two modules reduces the loss of the model in missing modality scenarios. 
Specifically, on the Baby data, when no modules are added, the performance loss is 37.15\%. 
After adding CEIM, the performance loss is 27.68\%; 
The performance loss after adding CMAM is 32.06\%. 
When the two modules cooperate, the performance loss is reduced to 24.59\%.
This shows that our model not only improves the performance of the model but also alleviates the performance degradation and enhances the generalization ability of the model.

 \begin{figure} [ht]
  \begin{center}
      \subfigure[$\alpha$ on Baby]{
      \includegraphics[width=0.15\textwidth]{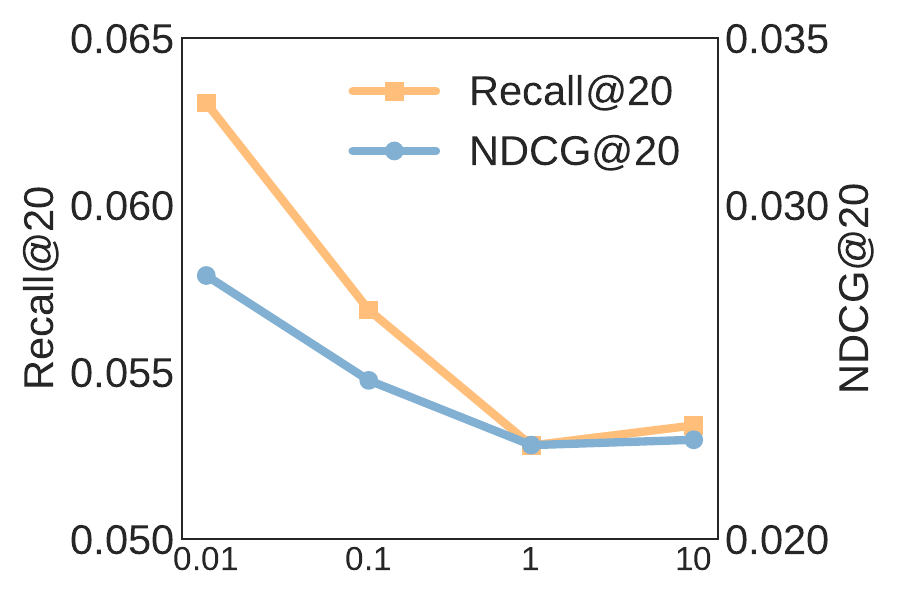}} 
      \subfigure[$\alpha$ on Clothing]{
      \includegraphics[width=0.15\textwidth]{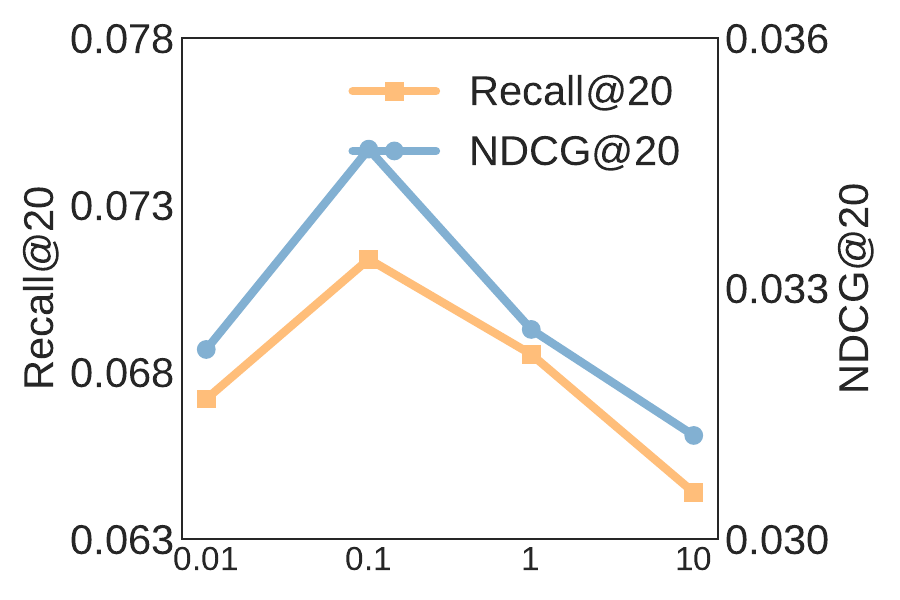}}
      \subfigure[$\beta$ on Baby]{
      \includegraphics[width=0.15\textwidth]{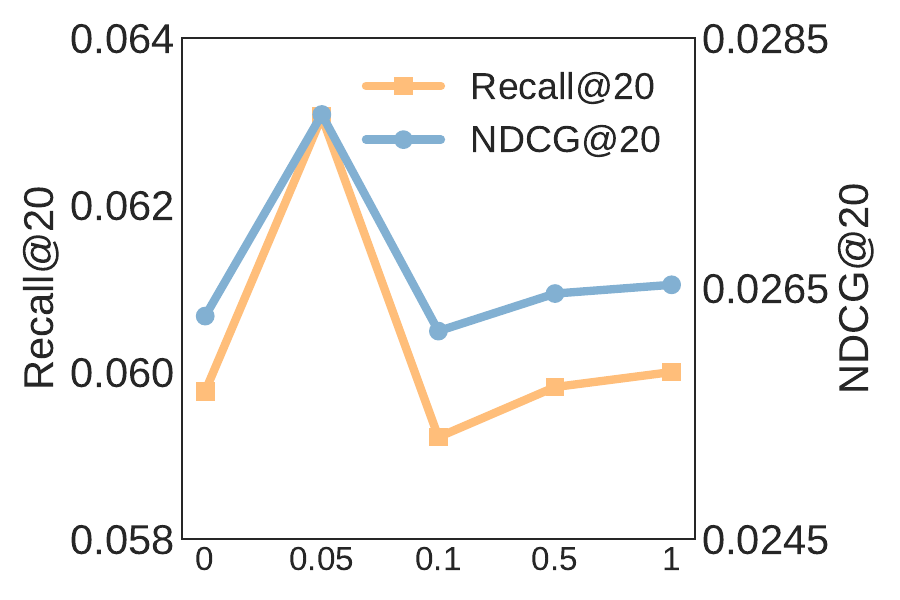}} 
      \\
      \subfigure[$\beta$ on Clothing]{
      \includegraphics[width=0.15\textwidth]{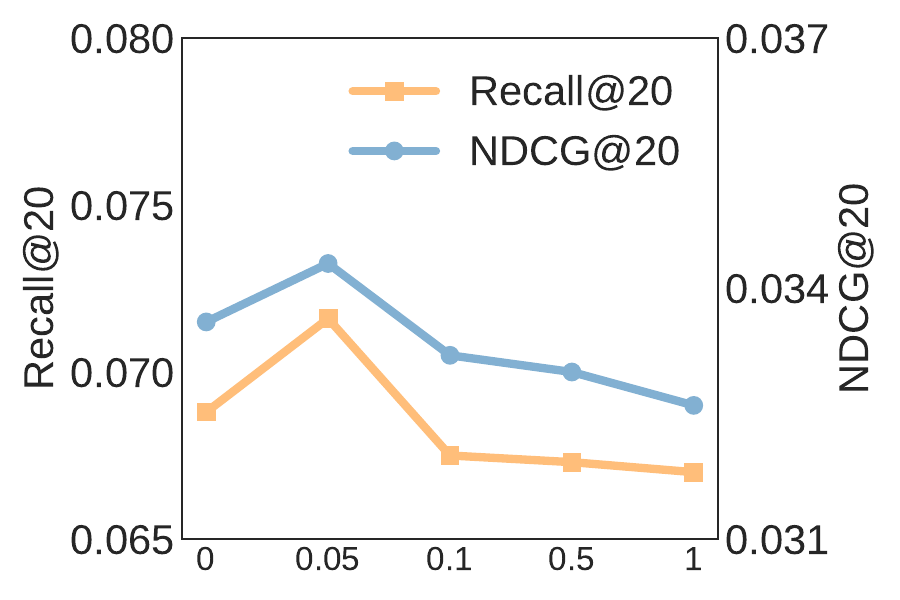}}
      \subfigure[ $\lambda$ on Baby]{
      \includegraphics[width=0.15\textwidth]{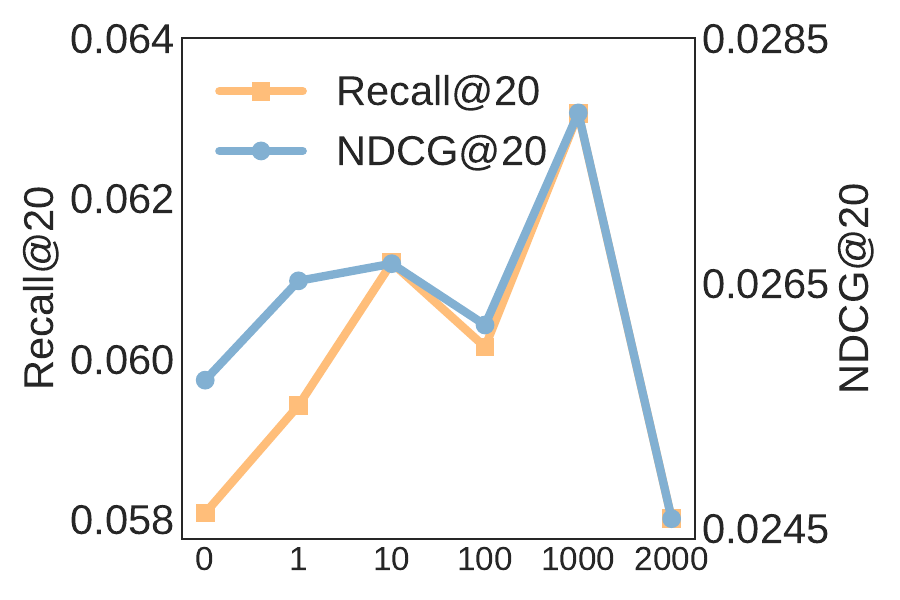}}
      \subfigure[ $\lambda$ on Clothing]{
      \includegraphics[width=0.15\textwidth]{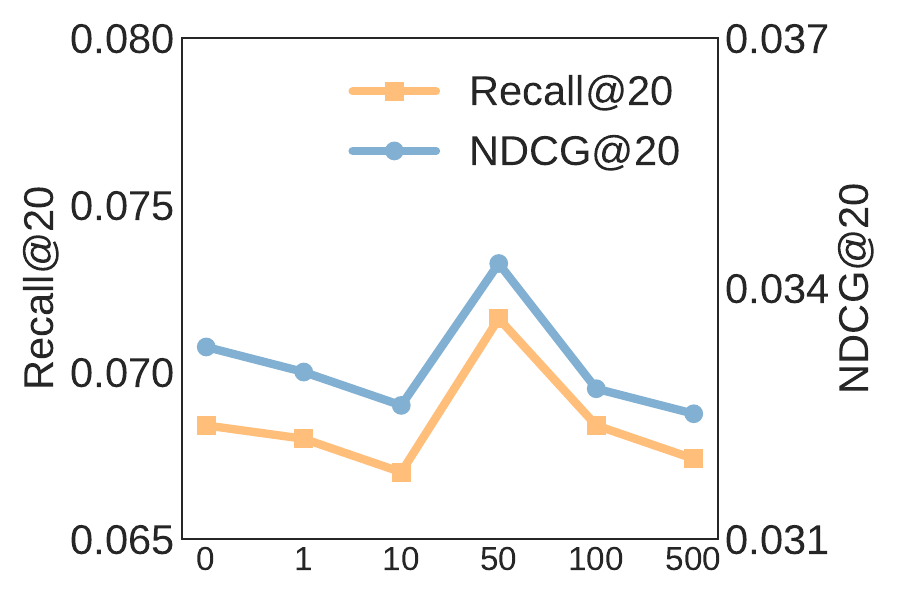}}
  \end{center}
  \caption{\small{Performance of different hyperparameters.}} 
  \label{fig: sensitivities}
\end{figure}

\subsubsection{Hyper-Parameter Sensitivities} \label{para} \hspace*{\fill}

\noindent\textbf{Effect of Dirichlet Distribution Parameters $\alpha$.}
As illustrated in Figure \ref{fig: sensitivities}(a) and (b), we sample weights from 4 different Dirichlet distributions and observe the performance of the model. Simply put, the smaller $\alpha$ is, the larger the weight gap is likely to be. In the FTMT setting, the model achieves the best results on the Baby and Clothing datasets with $\alpha=0.01$ and $\alpha=0.1$, respectively. On both datasets, the optimal value is achieved when $\alpha$ is small, which is due to the fact that one of the weights generated by Dirichlet distribution at this time will be around 1, which guarantees that the heterogeneous environment is dominated by a certain modality. Small $\alpha$ value helps ~\shortname~ to ensure that the modal composition in different environments is sufficiently different.

\noindent\textbf{Effect of Cross-Modality Alignment Loss Weights $\beta$.}
As illustrated in Figure \ref{fig: sensitivities}(c) and (d), we carefully tune the Cross-Modality alignment loss weights $\beta$ on the Baby and Clothing datasets.
We observe that \shortname~ achieves the best performance when $\beta=0.05$ on both datasets. 
When the weight is too small, the goal of capturing the information of other modalities cannot be achieved. When the weight is too large, the information of the modality itself may be seriously offset, resulting in performance degradation.

\noindent\textbf{Effect of Cross-Environment Invariant Loss Weights $\lambda$.}
As illustrated in Figure \ref{fig: sensitivities}(e) and Figure \ref{fig: sensitivities}(f), we carefully tune the cross-environment consistency loss weights $\lambda$ on the Baby and Clothing datasets.
\shortname~ achieves the best performance when $\lambda=1000$ on both the Baby and $\lambda=50$ on the Clothing datasets. The hyperparameter $\lambda$ has a direct impact on the model. Too small $\lambda$ results in cross-environment invariants where the invariant constraint is not sufficient to achieve our goal. Too large $\lambda$ will cause the model to focus too much on consistency and not enough on learning the recommendation task itself. In the implementation process, it needs to be carefully tuned to find the optimum.

\section{Conclusion}
In this paper, we focused on the problem of new item recommendations in the missing modality scenario.
We pointed out that existing works have too ideal assumptions about the data and are difficult to handle in real-world complex situations.
We proposed \shortname~ to solve the problem, encouraging the model to guarantee its performance as much as possible when the modality quality changes.
Specifically, we designed the cross-modal alignment module so that the single modal representation can capture the signals of other modalities.
Then we proposed a novel cyclic mixup method to construct multiple heterogeneous environments.
Based on these environments, we optimized our model using invariant learning, encouraging it to learn the users’ inherent content preferences that are kept invariant to arbitrary modality missing.
Extensive experiments verify the effectiveness of \shortname.

\newpage

\begin{acks}

This work was supported in part by grants from the National Key Research and Development Program of China(Grant No.2021ZD0111802), the National Natural Science Foundation of China(Grant No.U23B2031, 72188101), and the Fundamental Research Funds for the Central Universities (Grant No.JZ2023HGQA0471). 
\end{acks}
    
\bibliographystyle{ACM-Reference-Format}
\balance
\bibliography{References.bib}


\end{document}